\documentclass[12pt]{iopart}
\usepackage[utf8]{inputenc}
\usepackage{hyperref}
\usepackage{amsfonts}
\usepackage{amssymb}
\usepackage{graphicx}
\usepackage{siunitx}
\newcommand{\be}{\begin{equation}}
\newcommand{\ee}{\end{equation}}
\newcommand{\bea}{\begin{eqnarray}}
\newcommand{\eea}{\end{eqnarray}}

\newcommand{\bx}{{\bf x}}

\newcommand{\ket}[1]{\left|#1\right>}
\newcommand{\bra}[1]{\left<#1\right|}

\newcommand{\nne}{\nonumber \\ & = &}

\newcommand{\nnt}{\nonumber \\ & & \times}

\newcommand{\var}{{\rm var}}
\newcommand{\cov}{{\rm cov}}

\newcommand{\supi}{^{(i)}}

\newcommand{\supij}{^{(ij)}}
\newcommand{\subij}{_{ij}}

\graphicspath{{./}}

\usepackage{color}
\definecolor{mygreen}{rgb}{0,0.5,0} 
\definecolor{mygrey}{rgb}{0.5,0.5,0.5} 
\definecolor{myred}{rgb}{0.75,0,0} 
\definecolor{myblue}{rgb}{0,0,0.75} 
\definecolor{myviolet}{rgb}{0.5,0,0.5} 
\definecolor{mymagenta}{cmyk}{0,1,0,0.12} 
\definecolor{mycyan}{cmyk}{1,0,0,0.12} 
\definecolor{myorange}{rgb}{1,0.5,0}

\newcommand{\xset}{{\{\bx_i\}}}
\newcommand{\sset}{{\{\bss_i\}}}
\renewcommand{\vec}[1]{\mathbf{#1}}

\renewcommand{\S}{\hat{S}}

\renewcommand{\var}{\mathrm{var} }

\newcommand{\bS}{\vec{S}}
\newcommand{\bss}{{\bf s}}

\newcommand{\Hdd}{H_{\rm dd}}

\newcommand{\bn}{{\bf n}}
\newcommand{\dc}{{dc}}
\newcommand{\rf}{{rf}}

\usepackage[resetlabels]{multibib}
\newcites{SI,ME}{nada,nada}

\begin{document}
 
\newcommand{\mytitle}{Scale-invariant spin dynamics and the quantum limits of field sensing} 
\title{\mytitle
 }

\author{Morgan W. Mitchell}
\address{ICFO-Institut de Ciencies Fotoniques, The Barcelona Institute of Science and Technology, 08860 Castelldefels (Barcelona), Spain}
\address{ICREA -- Instituci\'o Catalana de Recerca i Estudis Avan\c{c}ats, 08010 Barcelona, Spain}
\ead{morgan.mitchell@icfo.eu}
\vspace{10pt}
\begin{indented}
\item[]\date{\today}
\end{indented}

\begin{abstract}
We describe quantum limits to field sensing that relate noise, geometry and measurement duration to fundamental constants, with no reference to particle number. We cast the Tesche and Clarke (TC) bound on dc-SQUID sensitivity as such a limit, and find analogous limits for volumetric spin-precession magnetometers.  We describe how randomly-arrayed spins, coupled to an external magnetic field of interest and to each other by the magnetic dipole-dipole interaction, execute a spin dynamics that depolarizes the spin ensemble even in the absence of coupling to an external reservoir. We show the resulting spin dynamics are scale invariant, with a depolarization rate proportional to spin number density and thus a number-independent quantum limit on the energy resolution per bandwidth $E_R$.  Numerically, we find $E_R \ge \alpha \hbar$, $\alpha \sim 1$, in agreement with the TC limit, for paradigmatic spin-based measurements of static and oscillating magnetic fields. 
\end{abstract}
\maketitle

\newcommand{\BW}{\Delta \nu}
\newcommand{\Btrue}{B_{\rm true}}
\newcommand{\szi}{\tilde{s}_{z,i}}
\newcommand{\szj}{\tilde{s}_{z,j}}
\newcommand{\supz}{^{(z)}}
\newcommand{\supjk}{^{(jk)}}
\newcommand{\supik}{^{(ik)}}
\newcommand{\supk}{^{(k)}}

\newcommand{\rhat}{{\bf  r}}
\newcommand{\bdir}{{\bf b}}
\renewcommand{\rhat}{\pmb{\scriptr}}
\renewcommand{\rhat}{{\mathbb R}}
\newcommand{\zhat}{{\mathbb Z}}
\newcommand{\Slim}{S_{\rm lim}}
\newcommand{\by}{{\bf y}}
\newcommand{\omdd}{\omega_{\rm dd}}
\renewcommand{\omdd}{\Upsilon_{\rm dd}}
\newcommand{\omL}{\omega_{L}}

The quantum limits of measurement is a rich topic of both fundamental and practical interest.   The theory of these limits informs many other topics, including the statistics of parameter estimation \cite{Holevo1982},  the geometry of quantum states \cite{BraunsteinPRL1994}, entanglement in many-body systems \cite{SorensenPRL2001}, quantum information processing \cite{LeeJMO2002,GiovannettiPRL2006}, and quantum non-locality \cite{TuraS2014,SchmiedS2016}.  Understanding quantum measurement effects has led to improved sensitivity in gravitational wave detectors \cite{LIGONP2011, AasiNP2013Short, LIGOCQG2015} and progress toward similar improvements in measurements of time \cite{AppelPNAS2009, LerouxPRL2010, LouchetChauvetNJP2010, ChenPRL2011, HostenN2016}, dc magnetic fields  \cite{WolfgrammPRL2010, SewellPRL2012} and radio-frequency fields \cite{WasilewskiPRL2010, MartinPRL2017}.

The vast majority of prior work on quantum sensitivity limits concerns the problem of linear interferometric parameter estimation.  For example, the standard quantum limit (SQL) $\langle\delta\phi^2\rangle \ge 1/N$ and the Heisenberg limit {(HL)} $\langle\delta\phi^2\rangle \ge 1/N^2$ constrain linear estimation of a phase $\phi$ given the resource of $N$ non-interacting two-level systems.  These dimensionless limits acquire units, e.g. length or time, through implementation-dependent scale factors, e.g. a wavelength or a transition frequency.  Because these scale factors, as well as the available $N$, can vary greatly from one implementation to another, {such dimensionless limits} do not by themselves provide {benchmarks} by which to compare different sensor implementations.

\newcommand{\limsym}{{\cal S}}
\newcommand{\name}{{noise power spectral density}}
\newcommand{\acr}{{NPSD}}

Here we study a qualitatively different kind of quantum sensitivity limit, one that contains no implementation-specific scale factors, and no makes reference to available resources, only to the quantity to be measured and to the method of measurement. To see what form such a limit could take, consider a sensor that measures the field $B$ in a volume $V$ over an observation time $T$, and gives a reading $B_{\rm obs} = B_{\rm true} + \delta B$, where $B_{\rm true}$ is the true value of the field and $\delta B$ is the measurement error -- a zero-mean random variable if the sensor is properly calibrated. The mean apparent magnetostatic energy in the sensor volume is $E_{\rm obs} = \langle B^2_{\rm obs}\rangle V/(2 \mu_0) =  B_{\rm true}^2 V/(2 \mu_0) +  \langle \delta B^2\rangle V/(2 \mu_0)$, where the second term expresses the sensor's so-called  ``energy resolution,'' which more properly can be identified as the bias of the apparent energy. 

Allowing for averaging of independent measurements of duration $T$, $\langle \delta B^2\rangle T$ is a figure of merit (smaller is better) that gauges the remaining error after unit total acquisition time. Allowing also for averaging over spatial region, $\langle \delta B^2\rangle VT$ is the relevant figure of merit. It is traditional to express this via the ``energy resolution per bandwidth'' 
\begin{equation}
\label{eq:ERL}
E_R \equiv \frac{\langle \delta B^2\rangle VT } {2 \mu_0} = \frac{S_B(0) V } {2 \mu_0} 
\end{equation}
where $S_x(0)$ is the low-frequency limit of the power spectral density of variable $x$. We seek a limit of the form $E_R  \ge \limsym$, where $\limsym$  is a constant with units of action.  As a purely empirical observation, several technologies to sense low-frequency magnetic fields approach $E_R = \hbar$, while to date none surpasses it \cite{MitchellRMP2020}. 
Such a limit makes no reference to available resources, and the scale factors are the fundamental constants $\mu_0 $ and $\hbar$. This describes a quantum limit resembling in some ways the relation $\Delta E_x \Delta B_y \ge 2 \pi \hbar c V^{-4/3}$ 
derived by W. Heisenberg \cite[pp. 48--54]{HeisenbergBook2013} where $c$ is the speed of light and $\Delta E_x$ and $\Delta B_y$ are uncertainties of orthogonal components of the electric and magnetic field, respectively. Whereas Heisenberg's result bounds the joint uncertainty of a pair of observables and allows $\Delta B = 0$ if the electric field $E$ is completely uncertain, our results  bound the magnetic field uncertainty while making no reference to the electric field.  

We can obtain a first implementation-independent limit from a well-known analysis of dc superconducting quantum interference devices (dc SQUIDs)  by Tesche and Clarke (TC) \cite{TescheJLTP1977}. Considering a lumped-circuit model for dc SQUID magnetometers with resistively-shunted Josephson junctions, TC computed the sensitivity, i.e. power spectral density of the equivalent noise, for an optimized device. At zero temperature, the sensitivity is limited by zero-point current fluctuations in the shunt resistances, to give ${S_\Phi(0)}/({2 L}) \ge \hbar$
where  $\Phi$ is the flux through the SQUID loop, and $L$ is the loop inductance  \cite{KochAPL1981, RobbesSAA2006}. 
With careful construction, small dc SQUID devices have reported ${S_\Phi(0)}/({2 L})$ as low as $2 \hbar$
\cite{AwschalomAPL1988, WakaiAPL1988, MuckAPL2001}. 

The implementation-dependent factor $L$ can be eliminated by noting that a wire loop has $\Phi = B A$ and $L =  \sqrt{A} \mu_0/\alpha$, where $A$ is the loop area and $\alpha$ is a wire-geometry factor of order unity\footnote{Any accurate discussion of the magnetostatics of wire loops leads into difficult geometrical problems that do not much concern us here. For this to be an implementation-independent limit, it suffices that $A/L^2$, and thus $\alpha$, is bounded from below.}.   This gives  
a limit that concerns only {field geometry and time}:  When measuring the field on a patch of area $A$ in a time $T$ with a dc-SQUID, the limiting sensitivity is equivalent to minimum energy per bandwidth of  
$\langle \delta B^2\rangle  A^{3/2}T/ (2 \mu_0 ) \ge  \alpha \hbar$.

In what follows we illustrate and derive an analogous limit for a second important field sensing technology, the spin-precession sensor. As the name suggests, such devices detect magnetic fields by the precession induced in an ensemble of spins.  Notable examples include alkali vapors \cite{Dang2010, GriffithOE2010}, nitrogen-vacancy centers in diamond \cite{WolfPRX2015, LovchinskyS2016}, and spinor Bose-Einstein condensates \cite{Vengalattore2007, PalaciosNJP2018}.  One might expect such sensors to be described by the ``quantum metrology'' analysis of linear interferometry \cite{GiovannettiPRL2006}, which gives rise to the SQL and HL given above. If this were the case, there could be no energy resolution limit, because the spin number density $\rho = N/V$ could be taken to infinity, such that $1/N$ and thus $\langle \delta B^2 \rangle$ approach zero for fixed $V$, leaving vanishing $E_R$. At high densities interactions cannot be neglected, however, and the linear interferometry results are not directly applicable \cite{MitchellQST2017}.

Prior works on quantum sensing with interacting particles have considered scaling \cite{BoixoPRL2007, NapolitanoN2011} and optimization \cite{MitchellQST2017, LuciveroPRA2017} scenarios. Here we show how  the mutual magnetic interaction of spins in a sensing ensemble can produce a self-similar spin dynamics resulting in implementation-independent energy resolution. Numerically, we find also ${\cal S} = \alpha \hbar$, in agreement with the TC analysis. Prior modeling of high-density NV-center ensembles  \cite{TaylorNP2008} and high-density alkali vapors \cite{KominisN2003,JimenezBook2017} have noted invariance of sensitivity with respect to spin density.

\newcommand{\psinaught}{| \psi(0) \rangle}
\newcommand{\unkn}{{\cal B}}
\newcommand{\Topt}{T_{\rm opt}}

We consider a generic prepare-evolve-project sensing protocol, in which an ensemble of $N$ spins $\sset$, each with spin quantum number $s$, is initialized in a product state $ | \Psi_0\rangle =  | \psi_0 \rangle^{\otimes N}$, allowed to evolve under a Hamiltonian $H$ containing as parameters the unknown positions of the spins $\xset$ and a field component $\unkn$  to be estimated, e.g.  the dc or rf field amplitude. The spins are detected {at time $T$} by projection of the total spin $\bS \equiv \sum_i {\bf s}_i$ onto a direction $\bn$.  

This model describes an idealized spin-precession sensor, without real-world complications such as the four possible spatial orientations of NV centers in a diamond lattice.  We focus on magnetic dipole-dipole interactions among the sensing spins as a source of spin depolarization. In contrast to the much studied effects of a nuclear spin bath \cite{YangRPP2016}, dipolar coupling among the sensing spins is intrinsic to the sensor rather than a feature of the environment \cite{BalasubramanianNMat2009}, and will dominate at high densities \cite{ZhouARX2019}.  We also note that some proposed spin-precession sensors employ other many-body effects including ferromagnetism \cite{JacksonKimballPRL2016} and quantum degeneracy \cite{PalaciosNJP2018} that are not included here. 

\begin{figure*}[t]
 \includegraphics[width=0.39 \textwidth]{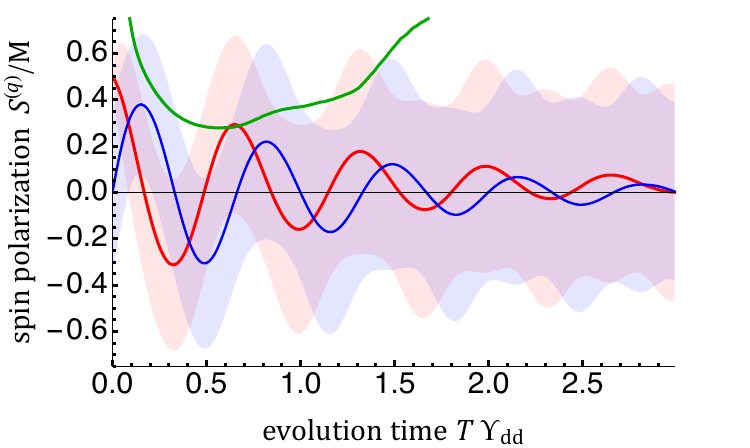}  \hspace{-18mm}
 \includegraphics[width=0.39 \textwidth]{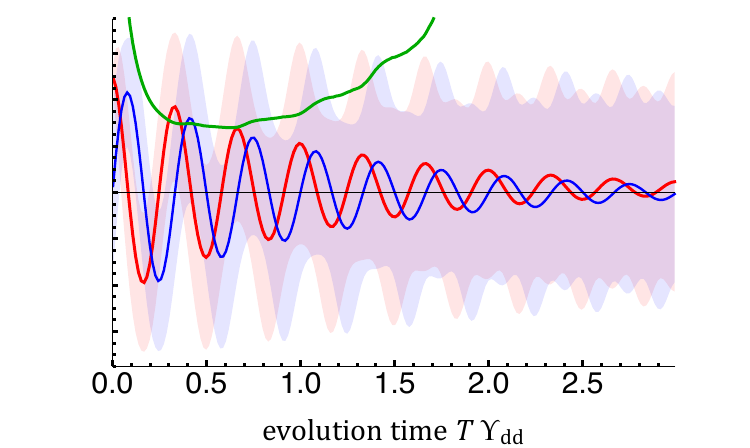}  \hspace{-18mm}
 \includegraphics[width=0.39 \textwidth]{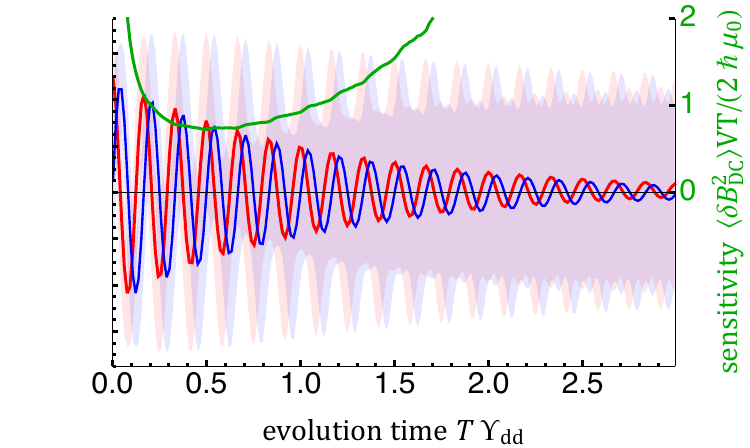}
\caption{Time evolution of spin statistics and sensitivity for fixed-spin dc magnetometry for a variety of magnetic moments and field strengths, showing the general behaviour of {seemingly} exponential loss of coherence through mutual interaction of the spins in the ensemble, and sensitivity nearly independent of implementation specifics. Following an initial tip into the plane orthogonal to the magnetic field, an ensemble spin-1/2 particles, with fixed random positions, is allowed to freely evolve.  Red and blue graphics show, on the left scale, the per-spin polarization $S^{(q)}_x/M$ and $S^{(q)}_y/M$, respectively, with curves showing mean value and shaded regions showing mean plus/minus {one rms} deviation, i.e. square root of the corresponding diagonal elements of the spin covariance matrix $\Gamma_S$.  Green curve shows, on the right scale, $E_R/\hbar$ with a minimum of $\approx 0.7$ at $T = T_{\rm opt} \approx 0.5 \omdd^{-1}$.  Computed using an ensemble of $Q=\SI{4e4}{}$ clusters of $M=2$ PPP-distributed spins.  From left to right: $(\omdd,\omega_L/2\pi) = (2,3), (1,3)$ and $(1,6)$.}
 \label{fig:illustration}
\end{figure*}

 Taking $V$ as constant, $T  \langle \delta \unkn^2 \rangle$ determines the energy resolution.  For  large $N$,  $\bS \cdot \bn$ is nearly  gaussian, and optimal estimation \cite{Helstrom1976} of $\unkn$ can be understood using propagation of error: If $\Gamma_S$ is the covariance matrix of $\bS$, with elements $\langle S_iS_j + S_j S_i\rangle/2 - \langle S_i\rangle \langle S_j \rangle$, 
the sensitivity limit is (see \ref{app:OptimalReadout})
\bea 
\label{eq:Sensitivity}
 T \langle\delta\unkn^2\rangle &\ge &   \min_{{\bf n},T}  T \frac{\bn \cdot \Gamma_S(T) \cdot \bn }{\left|{}{\partial_{\unkn}}  \langle \bS \cdot {\bf n}  \rangle\right|^{2}  }.
\eea
Here the expectation is defined as
\bea 
\label{eq:Expectation}
\langle {\cal A} \rangle &=& \int   \langle \Psi_0| U^{\dagger}(t) {\cal A} U(t)  |\Psi_0 \rangle P_\rho(\xset) \, d \xset, 
\eea
and includes classical averaging over configurations $\xset$ with probability density $P_\rho( \{ \bx \} )$, parametrized by $\rho$.  For example, $\xset$ could be distributed as a Poisson point process (PPP), in which an infinitesimal volume $dV$ contains a spin with probability $\rho\, dV$.   $U$ is the solution to {the Schr\"{o}dinger equation} and thus depends on $\xset$ and $\unkn$. 
 Due to coherent signal accumulation at short times and decoherence at long times, the minimum in Eq.~(\ref{eq:Sensitivity}) occurs at a finite time $T = \Topt$.  Examples (explained in detail below) are shown in Fig.~\ref{fig:illustration}. 
  
Considering a uniform magnetic field ${\bf B}$ with a constant component $B_{\rm \dc}$ along the $z$-axis, and including the MDDI interaction (indicated with the subscript $_{\rm dd}$), the Hamiltonian is  \cite[p. 103]{Abragam}
\bea
\label{eq:Hlab}
H &=& H_L + \Hdd \\ 
H_L & \equiv & - \gamma \hbar \sum_i {\bf s}_i \cdot {\bf B} \\
\Hdd & \equiv & \sum_{i\ne j}   \frac{\gamma^2 \hbar^2 \mu_0}{4 \pi r_{ij}^3}
\left[ {\bf s}_i \cdot {\bf s}_j - 3({\bf s}_i\cdot\rhat\subij)({\bf s}_j \cdot \rhat\subij) \right]
\nne
 \hbar \sum_{i\ne j}  \frac{\omdd}{s^2 \rho r_{ij}^3}
\left[ {\bf s}_i \cdot {\bf s}_j - 3({\bf s}_i\cdot\rhat\subij)({\bf s}_j \cdot \rhat\subij) \right],
\eea
where  $\gamma$ is the gyromagnetic ratio, $\rhat\subij = {\bf r}_{ij}/r_{ij}$, ${\bf r}_{ij} \equiv {\bf x}_i - {\bf x}_j$ and $r_{ij} \equiv |{\bf r}_{ij}|$.   $\omdd \equiv {s^2 \gamma^2 \hbar \mu_0 \rho}/(4\pi){}$ is the strength of the dipole-dipole coupling expressed as an angular frequency.

 
{We now explore the general features of this problem through numerical simulation.}   Due to the rapid fall-off of $\Hdd\supij$ with $r_{ij}$, and the fact that $\Hdd\supij$ vanishes when averaged over a sphere of constant $r_{ij}$, the dynamics of any given spin ${\bf s}_i$ will be determined mostly by its nearest neighbors and by ${\bf B}$.  This motivates the following approximation: {we consider the full system of $N$ spins as an ensemble of $Q=N/M$ clusters of $M$ spins each, with each cluster evolving independently under $H$.  The collective spin is then $\bS = \sum_{q=1}^Q \bS^{(q)}$, where $\bS^{(q)} = \sum_{{\bf s}_i \in c^{(q)}} {\bf s}_i$ is the total spin of cluster $c^{(q)}$.  Within each cluster, positions $\{\bx_i^{(q)}\}$ are assigned by finding the $M-1$ closest, PPP-distributed points to $\bx_1^{(q)}$, which is taken as the origin.}  For a product-state initial condition, the $\bS^{(q)}$ are independent, so that $\Gamma_{S} = \sum_q \Gamma_{S^{(q)}}$.  We compute $U(T)$ {using Eq.~(\ref{eq:Hlab}) and matrix exponentiation for $Q \sim 10^4$ clusters, to find} $\Gamma_{S^{(q)}}(T)$, $\langle \bS^{(q)}(T)\rangle$, and its derivatives. 

 \begin{figure}[t]
 \includegraphics[width=0.55 \columnwidth]{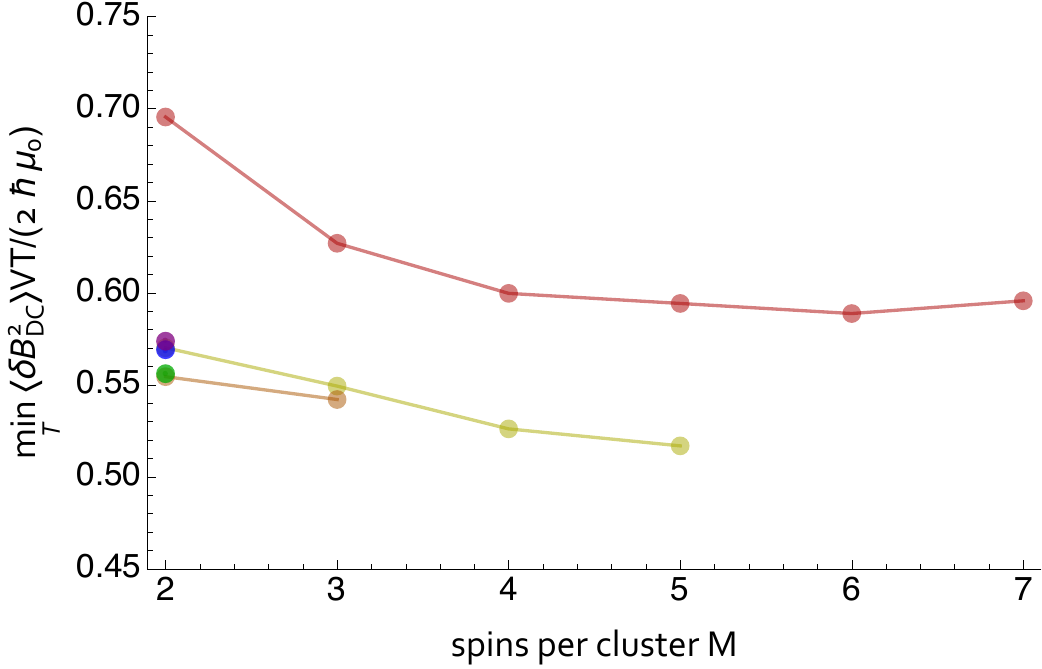}
 \caption{Convergence of numerical results with increasing number of spins $M$ and spin quantum number $s$.  Vertical axis shows time-optimal energy resolution, found as the minimum of $\langle\delta B_{\rm \dc}^2\rangle VT/(2\mu_0\hbar)$ versus $T$ as shown in \autoref{fig:illustration}. {Red, yellow, orange, green, blue and violet show $s=1/2, 1, 3/2, \ldots, 3$, respectively}. Computed with $Q=\SI{4e4}{}$.  }
 \label{fig:Convergence}
\end{figure}

We first consider the case of \dc~magnetometry, in which ${\bf B}(t) = (0,0,B_{\rm \dc})$, the initial state is $|\phi_0\rangle = |+x\rangle$, and  $\unkn = B_{\rm \dc}$.  
Representative results are shown in Fig.~\ref{fig:illustration}.  The mean amplitude of oscillation shows a steady and seemingly exponential decline, while the elements of the covariance matrix saturate, with the result that {the imprecision} reaches a minimum at which $ \langle\delta B_{\rm \dc}^2\rangle VT/(2\mu_0\hbar) \approx 0.7 $ at a finite time $T_{\rm opt}\omdd \approx 0.5$.   As shown in  Fig.~\ref{fig:Convergence}, the limiting sensitivity improves with increasing $M$, but saturates at about $M = 6$.  Similarly, $s$ larger than $1/2$ provides an advantage that appears to saturate about $s=1$.  Simulations (not shown) with other conditions, including different spin quantum number, gyromagnetic ratio, field strength and density,  find very similar limiting sensitivities, strongly suggesting an implementation-independent limit for fixed spin-precession sensors.  

\newcommand{\Uzero}{U_{L}} 
\newcommand{\UI}{U_{\rm dd}} 
\renewcommand{\UI}{U_{I}} 
\newcommand{\HI}{H_{I}} 

To understand {this limit}, we return to analytical methods.  
We shift to a coordinate system (for both spin orientations and positions) rotating about the field at the Larmor frequency, in effect working in the the interaction picture \cite{AulettaBook2009} with $\Hdd$ as interaction Hamiltonian.  We define rotating-frame coordinates and distances as $\tilde{\bf x}_i(t) = R_{z}(\omega_L t) {\bf x}_i(t)$ and $\tilde{\rhat}_{ij}(t) = R_{z}(\omega_L t) {\rhat}_{ij}(t)$, where the matrix $R_{z}(\theta)$ produces rotation about axis $z$ by angle $\theta$.  Rotating-frame spin operators $\tilde{\bf s}_i \equiv \Uzero(t) {\bf s}_i \Uzero^\dagger(t)$ are then defined via the unitary $\Uzero(t) \equiv \exp[i H_L t/\hbar]$, which is the inverse of time evolution under $H_L$ alone.  With this transformation, time evolution is governed by the rotating-frame Hamiltonian 
\bea
\label{eq:Hrot}
\tilde{H} &=& \Uzero(t) H \Uzero^\dagger(t) + i \dot{\Uzero}(t)\Uzero^\dagger(t) \nne
 \Uzero(t) H \Uzero^\dagger(t) - H_L \nne
 \hbar \sum_{i\ne j}  \frac{\omdd}{s^2 \rho r_{ij}^3}
\left[\tilde{\bf s}_i \cdot \tilde{\bf s}_j - 3(\tilde{\bf s}_i\cdot {\tilde\rhat}\subij)(\tilde{\bf s}_j \cdot {\tilde\rhat}\subij) \right].
\eea
The sensitivity limit given in Eq.~(\ref{eq:Sensitivity}) is unaffected by the frame shift, due to optimization over the read-out direction $\bn$.  In the rotating frame, the spins evolve only under their mutual coupling, which consists of two terms: the spin exchange term $\propto \tilde{\bf s}_i \cdot \tilde{\bf s}_j$ has no explicit time dependence and conserves total angular momentum, whereas the term  $\propto (\tilde{\bf s}_i\cdot {\tilde\rhat}\subij)(\tilde{\bf s}_j \cdot {\tilde\rhat}\subij)$ allows angular momentum to escape ``to the lattice'' and has coefficients (in ${\tilde\rhat}_{ij}$)  that oscillate at the Larmor frequency.

This periodicity of the rotating-frame Hamiltonian motivates a Kapitza approach in which we divide the dynamics of $\{ \tilde{\bf s}_i \}$ into a slowly-varying ``secular'' part and a ``micro-motion'' part oscillating at $\omega_L$. The micromotion is smaller than the secular part by a factor $\sim \omdd/\omL$, 
and for sufficient $\omL$ becomes negligible (see \ref{app:KTheory}). The secular motion is governed by the Larmor-cycle-averaged Hamiltonian
\bea
\label{eq:Hsec}
\bar{H}(t) &\equiv & \frac{1}{T_L} \int_{t}^{t+T_L} dt'\, \tilde{H}(t') 
\nne \hbar \sum_{i\ne j}      \frac{\omdd}{s^2 \rho r_{ij}^3}
\frac{1- 3 {\rhat}_{ij,z}^2}{2}  \left(3 \tilde{s}_{i,z}\tilde{s}_{j,z}    -  \tilde{\bf s}_i \cdot \tilde{\bf s}_j  \right),
\hspace{3mm}
\eea
where the subscript ${z}$ indicates the component along $\hat{z}$, i.e., along the dc field. The $x$ and $y$ components are lost in the cycle average.  
We note that $\omL$ no longer appears.

\newcommand{\xfo}{\mapsto}

We can now understand the effect of density in the secular regime,  using a strategy from renormalization group (RG) theory.  We imagine dividing the sensor volume into $\lambda$ equal sub-volumes, while also increasing the density by a factor $\lambda$. We indicate post-transformation quantities with primes, e.g. $\rho' = \lambda \rho$.   If  $P_\rho$ is self-similar, in the sense that the statistical distribution of $\{ \lambda^{1/3} ({\bf x}'_i -  {\bf x}'_j) \}$  within a sub-volume the same as that of $\{ {\bf x}_i -  {\bf x}_j \}$ in the full volume, and again assuming edge effects are negligible, the sub-volumes now represent $\lambda$ independent, reduced-scale realizations of the original sensor.  A PPP for example has such self-similarity. 

For a given configuration $\{ {\bf x}_i' \} = \{ \lambda^{-1/3} {\bf x}_i \}$, the Hamiltonian is $\bar{H}' =\lambda \bar{H}$, implying a speed-up of the rotating-frame dynamics  by a factor $\lambda$. When averaged over $\xset$, this produces faster evolution of spin means $\langle \tilde{\bss}'_i(t) \rangle = \langle \tilde{\bss}_i(\lambda^{} t) \rangle $ and correlators $\langle \tilde{\bss}'_i(t)  \tilde{\bss}'_j(t) \rangle = \langle \tilde{\bss}_i(\lambda t)  \tilde{\bss}_j(\lambda t) \rangle$.  Considering then the collective spin $\bS$, which sums the $\lambda$ sub-volumes, we find $\langle \tilde{S}'(t) \rangle = \lambda \langle \tilde{S}(\lambda^{} t) \rangle$ and $\Gamma_{S'}(t) = \lambda \Gamma_S(\lambda^{} t)$.   Inserting into Eq.~(\ref{eq:Sensitivity}), we find  $T_{\rm opt}' = \lambda^{-1} T_{\rm opt}$ and thus the same limiting sensitivity, independent of $\lambda$ and thus of $\rho$. 

The specific value of  $\gamma$ is similarly irrelevant:  a change to $\gamma' = \lambda \gamma$, through dynamics and quantum noise scaling, gives $\langle \tilde{S}'(t) \rangle = \lambda \langle \tilde{S}(\lambda^{} t) \rangle$, $\Gamma_{S'}(t) = \lambda \Gamma_S(\lambda^{} t)$, $T_{\rm opt}' = \lambda^{-1} T_{\rm opt}$, and again the same limiting sensitivity.  We see now that, within the secular regime $\omL\gg\omdd$, the only factors that can influence $E_R$ are $s$ and fundamental constants. This analytic result confirms the existence of an implementation-independent limit suggested by the numerical  calculations.

 \begin{figure}[t]
 \includegraphics[width=0.55 \columnwidth]{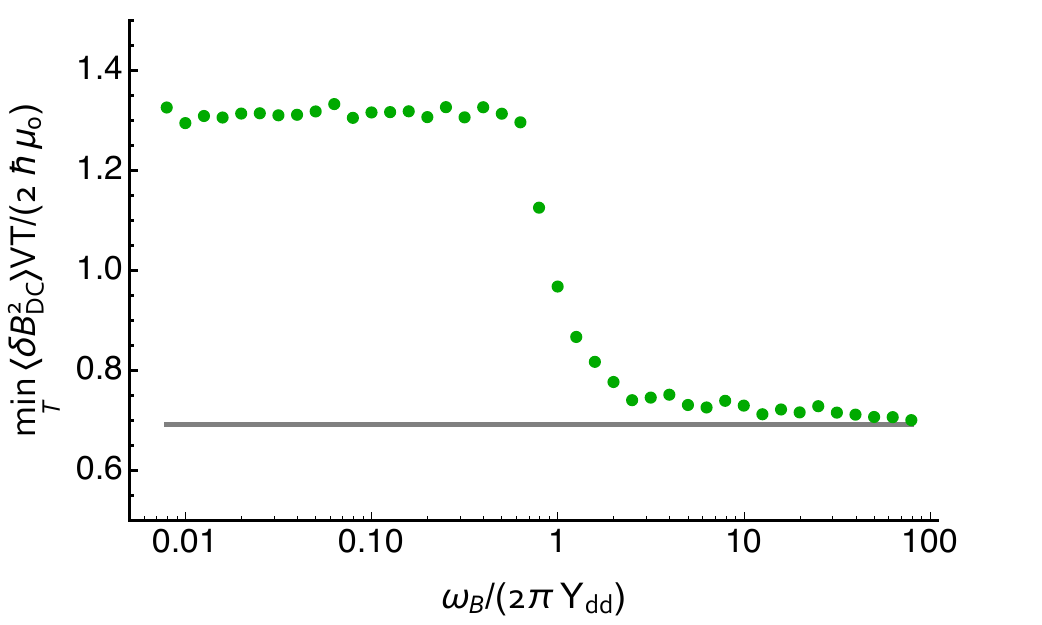}
 \caption{Transition from ``micro-motion'' to ``secular'' regime with increasing $\omL/\omdd$.  Green dots show  $\min_T \langle \delta B_{\rm \dc}^2 \rangle V T/(2\hbar\mu_0)$, i.e., $E_R/\hbar$, as obtained from simulations such as those shown in \autoref{fig:illustration}, using Eq.~(\ref{eq:Hlab}) to compute $U(t)$ and with $s=1/2$, $M=2$,  $Q=\SI{4e4}{}$.   Grey line shows the same quantity obtained using the secular approximation, i.e. using the $\omL$-independent Eq.~(\ref{eq:Hsec}) to compute $U(t)$.  }
 \label{fig:range}
 \end{figure}

Returning to numerical methods, we can show that the secular regime limit is in fact the global optimum. Again we use the cluster simulation. We study a range of $\omL / \omdd$, see Fig.~\ref{fig:range}, and find that outside of the secular regime $T\langle \delta B_{\rm \dc}^2 \rangle$  is about a factor of two larger than in the secular regime. 

\begin{figure}[t]
 \includegraphics[width=0.55 \columnwidth]{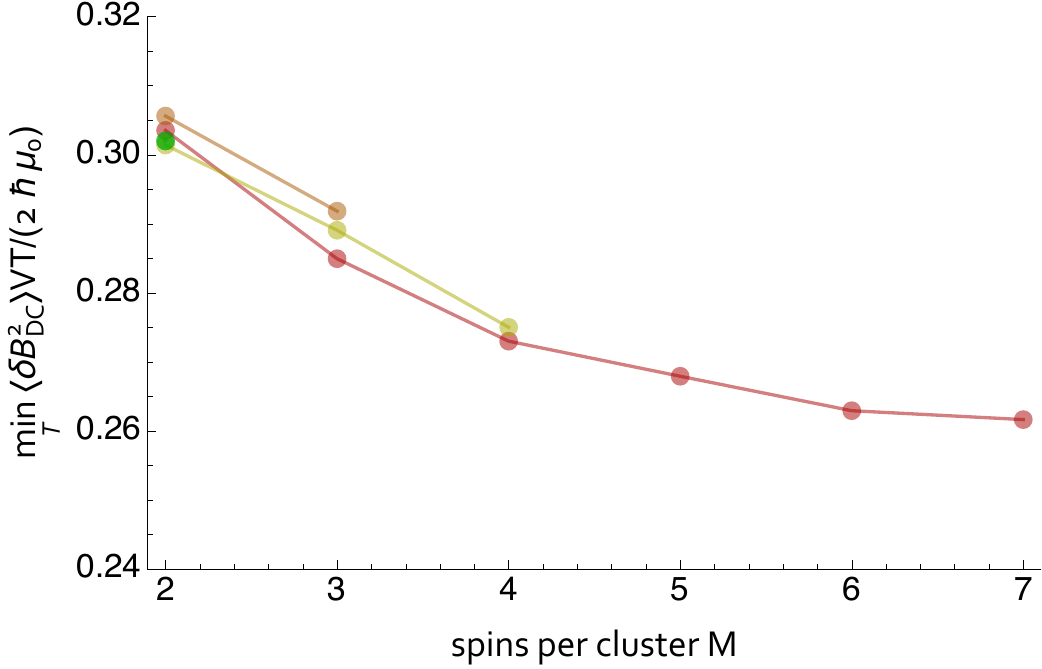}
\caption{RF sensitivity, as in Fig. \ref{fig:Convergence}.  Convergence of numerical results with increasing number of spins $M$ and spin quantum number $s$.  Vertical axis shows time-optimal energy resolution, found as the minimum of $\langle\delta B_{\rm \rf}^2\rangle VT/(2\mu_0\hbar)$ versus $T$ as shown in \autoref{fig:illustration}. {Red, yellow, orange, and green show $s=1/2, 1, 3/2$ and $2$, respectively}. Computed with $Q=\SI{1e4}{}$ except for $s=1/2$, $M=7$, computed with $Q=\SI{3e4}{}$.  }
 \label{fig:RFsensitivity}
\end{figure}

Using $\bar{H}$ we can also consider RF magnetometry \cite{MartinPRL2017}, which requires only minor modification to the above discussion. Now ${\bf B}(t) = (B_{\rm RF} \cos \omega t,B_{\rm RF} \sin \omega t,B_{\rm \dc})$,  the initial state is $|\phi_0\rangle = |+z\rangle$, and the unknown is $\unkn = B_{\rm RF}$.  As before, we 
take $\xset$ to be PPP-distributed.  In the rotating frame, the RF field appears fixed, and contributes a term $- \gamma \hbar B_{\rm RF} \sum_i \tilde{\bf s}_{i} \cdot \hat{x}$ to both Eqs.~(\ref{eq:Hrot}) and ~(\ref{eq:Hsec}). For  $\gamma |B_{\rm RF}| \ll \omdd$, i.e. for weak-field detection, the contribution of this term to the dynamics is small, and the scale-invariance arguments proceed as above. Here also, this shows analytically that there is an implementation-independent limit on the sensitivity.
Using Eq.~(\ref{eq:Hsec}) we numerically evaluate $\Gamma_{\tilde{S}}$ and $\partial_{B_{\rm RF}} \langle \tilde{\bf S} \cdot {\bf n} \rangle$ by the cluster expansion as above.  Results, shown in  Fig.~\ref{fig:RFsensitivity}, indicate a lower-bound of $E_R/\hbar \approx 1/4$.  

Thus far, the discussion has concerned only spins with fixed positions.  For mobile spins, we  consider a set of trajectories $\{ {\bf x}_i(t) \}$, and the RG argument proceeds as above if the trajectories' distribution is self-similar in the sense that $\{ \lambda^{1/3} [{\bf x}'_i(t) -  {\bf x}'_j(t)] \}$ has the same distribution as $\{ {\bf x}_i(\lambda t) -  {\bf x}_j(\lambda t) \}$.  This describes sub-diffusive $(\Delta x)^3 \propto \Delta t$  transport, as opposed to diffusive $(\Delta x)^2 \propto \Delta t$ or ballistic  $\Delta x \propto \Delta t$ transport. As such, we cannot directly use the RG argument to establish an implementation-independent limit for vapor- or gas-phase spin-precession sensors. Nonetheless, we recover scale-invariance in two scenarios.  

First, precisely because the decoherence rate $\omdd \propto \rho$ grows faster with density than does the $t \propto \rho^{2/3}$ (diffusive) or $t \propto \rho^{1/3}$ (ballistic)  transport time across the inter-particle spacing, for sufficiently large densities (or sufficiently slow transport) the spins will appear effectively immobile and can be treated as fixed, so that the conclusions given above for fixed spins apply.  Second, in the opposite extreme of highly-mobile, weakly-coupled spins, we can expect short-range collisional processes to cause spin depolarization faster than do long-range spin-spin interactions. Such 
collisions  produce a decoherence rate $\Gamma \propto \rho$ in both diffusive and ballistic regimes \cite{HapperBook2010}.  As a consequence, $E_R$ becomes independent of density $\rho$.  The collisional processes that produce decoherence 
do not, however, appear to scale with $\gamma$ as would be required for a species-independent sensitivity limit \cite{AllredPRL2002}.  For this regime, we thus expect a density-independent limit for any given sensing species, but  we do not expect these limits to be the same for different species.

\section*{Conclusions}
We have identified a new kind of quantum sensing limit, one that applies to dimensioned physical quantities such as length or field strength, but which makes reference neither to available quantum resources such as particle number, nor to implementation-dependent scale factors such as the sensing particles' wavelength or transition frequency.  For spin-precession sensors, the limit is a consequence of scale-invariance in the self-decoherence dynamics of spin-ensembles. {For sensors employing fixed, randomly-placed spins, the limiting ``energy resolution per bandwidth'' is near the reduced Planck constant, a result that coincides with the limit for dc-SQUID sensors.} 

Why such a limit would fall so close to the quantum of action is an intriguing open question. Regarding this, we make the following observations: Previous quantum sensing limits for open quantum systems \cite{HuelgaPRL1997, EscherNP2011, Demkowicz-DobrzanskiNC2012} have considered non-interacting spins that experience decoherence due to independent coupling to a reservoir. These models arrive to limits of the form $\langle \delta X^2 \rangle \propto 1/N$, where $X$ is a quantity to be sensed. The limit here is both more restrictive, in the sense that it does not allow arbitrarily good sensing in the $N\rightarrow \infty$ limit, and of a qualitatively different origin. Here the decoherence occurs not because the spins each interact with a reservoir, but rather because the sensing spins interact with each other. More precisely, the dipole-dipole coupling allows disorder in the centre-of-mass degrees of freedom to enter the spin degrees of freedom. As the new limit intrinsically involves an interacting many-body system, one might hope for insights from the theory of open quantum many-body systems.  To our knowledge the problem has not be discussed in that context.  We leave for future work the interesting question of whether the limit can be ``beaten'' by dynamical decoupling of interacting spins \cite{PhamPRB2012, ChoiARX2019, ZhouARX2019}, or by using spin ensembles with reduced centre-of-mass entropy,  e.g. spinor Bose-Einstein condensates \cite{PalaciosNJP2018} or microscopically-ordered spin systems. 

\ack
We thank J. Kitching, M. Lukin, H. Zhou, I. Chuang, S. Palacios and R. J. Sewell for helpful discussions.  This project was supported by the European Research Council (ERC) projects AQUMET (280169) and ERIDIAN (713682); European Union projects QUIC (Grant Agreement no.~641122) and FET Innovation Launchpad UVALITH (800901);~the Spanish MINECO projects OCARINA (Grant Ref. PGC2018-097056-B-I00) and Q-CLOCKS (PCI2018-092973), the Severo Ochoa programme (SEV-2015-0522); Ag\`{e}ncia de Gesti\'{o} d'Ajuts Universitaris i de Recerca (AGAUR) project (2017-SGR-1354); Fundaci\'{o} Privada Cellex and Generalitat de Catalunya (CERCA program, RIS3CAT project QuantumCAT); Quantum Technology Flagship projects MACQSIMAL (820393) and QRANGE (820405); Marie Sk{\l{odowska-Curie ITN ZULF-NMR (766402); 17FUN03-USOQS, which has received funding from the EMPIR programme co-financed by the Participating States and from the European Union's Horizon 2020 research and innovation programme. 

\appendix

\section{Optimal readout}
\label{app:OptimalReadout}
To efficiently evaluate Eq.~(\ref{eq:Sensitivity}), it is convenient to note that by parametrizing ${\bf n} = (\cos \theta, \sin \theta, 0)$, the variance of the estimate of $B$ can be written 
\bea
\langle \delta B^2 \rangle  &=&  \min_{\bf n}  \frac{\bn  \Gamma_S  \bn^T }{\left|{}{\partial_{\unkn}}  \langle \bS \cdot {\bf n}  \rangle\right|^{2}  } 
\nne  \min_\theta \frac{ (\cos \theta, \sin\theta, 0 )  \Gamma_S  (\cos \theta, \sin \theta, 0 )^T }{ \left[ \partial_\unkn ( \langle S_x \rangle \cos \theta +  \langle S_y \rangle \sin \theta ) \right]^{2} }
\nne \frac{|\Gamma|_2}{Z}
\eea
where 
\be
Z  \equiv  
 (\partial_{\unkn} \langle S_x\rangle)^2 \var S_y - 2 \cov(S_x,S_y) (\partial_\unkn  \langle  S_x\rangle) (\partial_\unkn   \langle  S_y\rangle ) + (\partial_\unkn  \langle  S_y\rangle)^2 \var S_x   
 \ee
and $| \cdot |_2$ indicates the determinant of the upper left $2\times 2$ sub-matrix.

\newcommand{\KStart}{t_\alpha}

\section{Kapitza-theory dynamics}
\label{app:KTheory}
To understand the conditions under which spin-interaction dynamics will simplify due to Larmor precession, we adapt the classical Kapitza method \cite{KapitzaBook1965, BandyopadhyayP2008} to the spin problem at hand. 
We begin by writing the equations of motion for the spins, in the frame rotating at $\omega_L$ and thus governed by Hamiltonian of Eq.~(6). We have 
 \bea
 \label{eq:RotatingFrameDynamics}
\frac{d}{dt} \tilde{\bf s} \supi &=& 
 \sum_{k\ne i}  \frac{\gamma^2 \hbar^2 \mu_0}{4 \pi r_{ik}^3}  
\left[  {3}
(\tilde{\bf s}\supi \times {\tilde\rhat}\supik) ({\tilde{\bf s}}\supk \cdot {\tilde\rhat}\supik)  +   ( \tilde{\bf s}\supi  \times  \tilde{\bf s}\supk) \right] 
 \eea
 where $\tilde{\bf r} \equiv R_z(\omega_L t) {\bf r}$ is the rotated vector joining the spins.  The precession period is $T\equiv 2 \pi / \omega_L$. 
 
Collecting all spin components into a single vector ${\bf  z} = \bigoplus_i  \tilde{\bf s} \supi $, we note that Eq.~(\ref{eq:RotatingFrameDynamics})  has the form
\be
\label{eq:AbstractRFD}
\frac{d}{dt} z_i  = {\cal R}_{ijk}(t) z_j z_k
\ee
where the tensor of coefficients ${\cal R}$ is periodic: ${\cal R}_{}(t+T) = {\cal R}_{}(t)$.  It is convenient to  identify a time $\KStart$ as the start of a cycle, and divide ${\cal R}_{}(t)$ as ${\cal R}(t) = \bar{\cal R}  + \mathring{\cal R}(t)$ where the cycle-averaged part is $\bar{\cal R} \equiv T^{-1} \int_{\KStart}^{\KStart+T} {\cal R}(t) dt$.  

 We write ${\bf z} = {\bf p} + {\bf q}$, where ${\bf p}$ is the slowly-varying ``secular motion'' and ${\bf q}$ is the small and rapidly-varying ``micro-motion,'' defined as the solution of
 \be
\frac{d}{dt} q_i  =  \mathring{\cal R}_{ijk}(t)
\ee
with zero cycle-average: $\int_{\KStart}^{\KStart +T} {\bf q}(t) dt = {\bf 0}$.  We can write the formal solution 
 \be
q_i(t)  =   \int_{\KStart}^{t} \mathring{\cal R}_{ijk}(t')  dt' \,\, p_j(\KStart) p_k(\KStart). 
\ee
We note that ${\bf q} \sim T$, and thus ${\bf q}$ becomes small for large $\omega_L$.

Using the smallness of  ${\bf q}$ we expand the r.h.s. of Eq.~(\ref{eq:AbstractRFD}), as applies to the time period $t \in [\KStart, \KStart+T)$  to find
\bea
\frac{d}{dt} z_i  &=&
{\cal R}_{ijk}(t) z_j(\KStart) z_k(\KStart) 
+ q_l(t) \left[\partial_{z_l}{\cal R}_{ijk}(t) z_j z_k\right]_{{\bf z} = {\bf p}(\KStart)} + O(q)^2.
\eea
We drop the  doubly-small $O(q)^2$ term and integrate over one cycle to find the cycle-averaged rate of change 
\bea
\label{eq:KAResult}
\frac{d p_i}{dt} \approx \frac{\Delta p_i}{T}  &=& \bar{\cal R}_{ijk}(t) p_j(\KStart) p_k(\KStart) 
+  \frac{1}{T} \int_{\KStart}^{\KStart+T} dt \int_{\KStart}^{t} dt' \mathring{\cal R}_{ijk}(t')   p_j(\KStart) p_k(\KStart) 
\nnt \left[\partial_{z_l}{\cal R}_{ijk}(t) z_j z_k\right]_{{\bf z} = {\bf p}(\KStart)},
\eea
which now refers only to ${\bf p}$.   The second term describes the effect of micromotion on the secular dynamics.  

We note that the first term in Eq.~(\ref{eq:KAResult}) scales as $\omdd$, and contains both the $\tilde{\bf s}\supi  \times  \tilde{\bf s}\supk$ factor that produces spin-exchange and $(\tilde{\bf s}\supi \times {\hat{z} \tilde\rhat}\supik_z) ({\tilde{\bf s}}\supk \cdot {\hat{z} \tilde\rhat}_z\supik)$ obtained by cycle-averaging  $(\tilde{\bf s}\supi \times {\tilde\rhat}\supik) ({\tilde{\bf s}}\supk \cdot {\tilde\rhat}\supik)$.  This latter factor is responsible for loss of angular momentum to the centre of mass degrees of freedom.  The second term in Eq.~(\ref{eq:KAResult}) scales as $\omdd^2/\omega_L$, i.e., smaller than the first by a factor $\omdd/\omega_L$.  It is this smallness that justifies using the cycle-averaged Hamiltonian of Eq.~(\ref{eq:Hsec}) for large $\omega_L$.

%
%
%

\renewcommand{\bra}[1]{\langle #1|}
\renewcommand{\ket}[1]{|#1\rangle}

\section*{References}
\bibliographystyle{./iopart-num}
\bibliography{./MegaBib}

\end{document}